\pdfoutput=1
\documentclass{article}

\usepackage[preprint]{neurips_2025}

\usepackage[utf8]{inputenc}
\usepackage[T1]{fontenc}
\usepackage{hyperref}
\usepackage{url}
\usepackage{booktabs}
\usepackage{amsmath}
\usepackage{graphicx}
\usepackage{xcolor}
\usepackage{tikz}
\usetikzlibrary{positioning,fit,calc,backgrounds}

\title{AKTS: Sub-Microsecond Kernel Policy Switching\\for Language-Model Agents}

\author{%
  Mohammadali Khodabandehlou\thanks{Both authors contributed equally.} \\
  University of Southern California \\
  \texttt{mk40089@usc.edu} \\
  \And
  Mahdi Alizadeh \\
  University of Southern California \\
  \texttt{malizade@usc.edu} \\
}

\begin{document}
\maketitle

\begin{abstract}
GPU-backed LLM servers multiplex interactive requests with background batch
work on the same CPUs; a fixed kernel policy serves one objective and loses the
other, so agentic OS control needs to switch scheduler behavior as the workload
changes. The hard part is applying the switch safely and fast enough for the
kernel: scheduler events occur every $1$--$10\,\mu$s, and any code running
there must satisfy the eBPF verifier. Scalar knobs are fast but limited, while
generating eBPF policy code puts compilation, verification and possible
rejection on the runtime path. We present AKTS, which verifies a policy library
\emph{once, at load time} and reduces the agent's runtime action to writing an
integer index into an in-kernel array of preverified policies, resolved by a
tail call. Because the agent emits an index rather than code, verifier failure
is not a runtime outcome. On Linux~6.14, AKTS applies a policy switch in
$920$\,ns (p50); makes an invalid index inert across $60{,}217$ invocations on
an attached scheduler; and switches policies in a vLLM workload to capture
$97\%$ of a throughput policy's batch work while matching a latency policy's
burst response.
\end{abstract}

\section{Introduction}

On a GPU-backed LLM serving node the CPUs still handle request intake,
tokenization, prefill orchestration and background jobs. During a burst those
CPUs should prioritize the serving path for low time-to-first-token; in quiet
periods they should let background work progress. A scheduler tuned for one
regime loses the other.

AKTS targets this problem: recognize the regime at a coarse timescale and
switch kernel scheduling behavior to match. The challenge is making that switch
a safe, cheap kernel operation. Schedulers place tasks every few microseconds
while even small models need milliseconds per decision, so inference on the
scheduling path is ruled out by construction. Prior work explores in-kernel
policy selection with eBPF~\cite{ebpfbandit} and adaptive scheduling for
dynamic workloads~\cite{dynamiciot,decima,ghost}; the missing piece is the
actuator, a safe, sub-microsecond mechanism letting a slow agent switch among
verified kernel policies. Agentic-OS work~\cite{infragym} has converged on two
ways around the timing boundary.

The first is \textbf{scalar tuning}. LumOS~\cite{lumos} has an LLM agent tune
Completely Fair Scheduler hyperparameters, outperforming Bayesian optimization
by $5$--$7\%$ and a human expert by $2.98\%$. Applying a decision is a single write to a knob the kernel already exposes, so
actuation is effectively free; the limitation is expressive, not temporal.

The second is \textbf{code synthesis}. SchedCP~\cite{schedcp} has an LLM
generate sched\_ext~\cite{schedext} scheduling policy code, which is then
compiled, checked by the in-kernel verifier, and loaded, reporting up to
$1.79\times$ improvement on target workloads. This recovers full expressiveness but places compilation and verification on
the actuation path, costing seconds to a minute per change, and generated code
can be rejected by the verifier and need regeneration. Learned schedulers~\cite{decima,firm,park} adapt policies but do not
synthesize mechanisms, and LLM-generated kernel extensions~\cite{kgent}
inherit the compile-and-verify path.

\textbf{AKTS} (Agentic Kernel Tail-Call Stitching) removes that tension. A
library of eBPF policies is compiled and verified once at load time, then
installed into a program array. At runtime, a dispatcher on the kernel hook
reads a single integer and tail-calls into the policy it names. The agent's
entire runtime output is that integer.

This yields a safety property neither prior approach has. The agent's action
space is an index into programs that were \emph{already verified}; there is no
code path by which a hallucinated decision becomes an unsafe kernel. Asked to
choose between policies $0$ and $1$, a $0.5$B model in our evaluation answered
\texttt{"0.3"}; under AKTS that output is inert, since an invalid choice names
no program at all. This complements LumOS's transactional
apply--commit--revert and SchedCP's pre-deployment analysis, which validate or
roll back a proposed action; AKTS makes unsafe actions inexpressible. Program
arrays and tail calls are long established in the networking
datapath~\cite{bpf}; our contribution is their use as a constrained action
space composing with sched\_ext.

We contribute: \textbf{(i)} the mechanism, which decouples policy
expressiveness from actuation cost by moving verification to load time
(\S\ref{sec:design}); \textbf{(ii)} \textbf{three feasibility results} on stock
Linux~6.14, each bracketed by controls, establishing that tail-call dispatch
verifies, populates and executes inside sched\_ext (\S\ref{sec:feasibility}); and
\textbf{(iii)} an evaluation showing actuation at parity with scalar tuning,
an invalid decision proven inert on an attached
scheduler, and a vLLM serving workload on which switching captures $97\%$ of a
throughput policy's batch work while matching a latency policy's burst
response (\S\ref{sec:eval}).

\section{Design}
\label{sec:design}

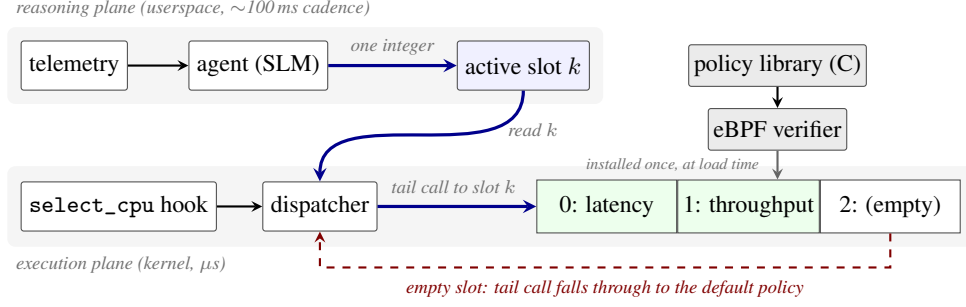
\begin{figure}[t]
\centering
\begin{tikzpicture}[
  font=\small,
  box/.style={draw=black!70, fill=white, rounded corners=1pt, minimum height=6.5mm, inner sep=3pt},
  slot/.style={draw=black!70, fill=white, minimum height=7mm, minimum width=18.5mm, inner sep=2pt},
  plane/.style={rounded corners=3pt, inner sep=5pt},
  lbl/.style={font=\scriptsize\itshape, text=black!55},
  role/.style={font=\scriptsize\itshape, text=black!45, align=center},
  arr/.style={-stealth, thick},
  agentarr/.style={-stealth, very thick, draw=blue!55!black},
  failarr/.style={-stealth, thick, dashed, draw=red!50!black}]

  \node[box] (telem) {telemetry};
  \node[box, right=8mm of telem] (agent) {agent (SLM)};
  \node[box, right=17mm of agent, fill=blue!6] (slotmap) {active slot $k$};
  \begin{scope}[on background layer]
  \node[plane, fill=black!4, fit=(telem)(agent)(slotmap),
        label={[lbl, anchor=south west]north west:reasoning plane (userspace, $\sim$100\,ms cadence)}] (rp) {};
  \end{scope}
  \draw[arr] (telem) -- (agent);
  \draw[agentarr] (agent) -- node[above, lbl] {one integer} (slotmap);

  \node[box, below=12mm of telem.south west, anchor=north west] (hook) {\texttt{select\_cpu} hook};
  \node[box, right=6mm of hook] (disp) {dispatcher};
  \node[slot, right=21mm of disp, fill=green!7] (s0) {0: latency};
  \node[slot, right=0mm of s0, fill=green!7] (s1) {1: throughput};
  \node[slot, right=0mm of s1] (s2) {2: (empty)};
  \begin{scope}[on background layer]
  \node[plane, fill=black!4, fit=(hook)(disp)(s0)(s2),
        label={[lbl, anchor=north west]south west:execution plane (kernel, $\mu$s)}] (ep) {};
  \end{scope}
  \draw[arr] (hook) -- (disp);
  \draw[agentarr] (disp) -- node[above=0.4mm, lbl, pos=0.48] {tail call to slot $k$} (s0);
  \draw[agentarr] (slotmap.south) to[out=-90,in=90] node[right=0.8mm, lbl, pos=0.25] {read $k$} (disp.north);
  \coordinate (fa) at ($(s2.south)+(0,-4.5mm)$);
  \coordinate (fb) at (disp.south |- fa);
  \draw[failarr] (s2.south) -- (fa) --
     node[below=0.4mm, lbl, text=red!50!black] {empty slot: tail call falls through to the default policy}
     (fb) -- (disp.south);

  \node[box, right=13mm of slotmap, fill=black!8, minimum height=5.5mm, inner sep=2pt] (lib) {policy library (C)};
  \node[box, below=2.8mm of lib, fill=black!8, minimum height=5.5mm, inner sep=2pt] (verif) {eBPF verifier};
  \draw[arr] (lib) -- (verif);
  \draw[arr, draw=black!60] (verif.south) -- (verif.south |- s2.north);
  \node[lbl, font=\tiny\itshape, anchor=east]
    at ($(verif.south)!0.5!(verif.south |- s2.north)+(-1.2mm,0)$) {installed once, at load time};

\end{tikzpicture}
\vspace{-1mm}
\caption{AKTS. Policies are compiled and verified once, at load time (gray).
The agent's entire runtime action is writing the integer $k$ (blue); the kernel
tail-calls into the selected preverified policy on every scheduling event. An
invalid $k$ names an empty slot and is inert (red).}
\label{fig:arch}
\end{figure}

AKTS separates a \emph{reasoning plane} running at the agent's cadence from an
\emph{execution plane} running at the kernel's (Figure~\ref{fig:arch}). All
kernel code is verified before the scheduler attaches; at runtime the agent can
only change the integer selecting a verified policy.

\paragraph{Execution plane.} At load time, each candidate scheduling policy is
compiled, verified, and installed into a \texttt{BPF\_MAP\_TYPE\_PROG\_ARRAY}.
A dispatcher on the target hook reads $k$ from a single-entry map and issues
\texttt{bpf\_tail\_call}, which transfers control without growing the stack
and uses one of 33 permitted tail calls per event. If $k$ is out of range or
names an empty slot, the dispatcher runs its default action.

\paragraph{Reasoning plane.} A small model reads telemetry every ${\sim}100$\,ms
and writes $k$ with \texttt{bpf\_map\_update\_elem}; between writes the kernel
dispatches with no model call or userspace round trip.

\paragraph{Safety.} The action space is \emph{closed}: the agent selects among
programs the verifier already accepted, so no output it can produce causes
unverified code to execute. An incorrect valid index can choose the wrong
policy; an invalid index chooses no policy. A designed but unimplemented
in-kernel guard can reset the slot without waiting for the agent.

\section{Feasibility on stock Linux}
\label{sec:feasibility}

The design rests on three undocumented assumptions about sched\_ext and tail
calls; it fails if any does not hold. All measurements in this paper use one host:
an NVIDIA A100-SXM4-40GB (40\,GB) with a 30-vCPU AMD EPYC 7J13 and 216\,GB RAM,
Ubuntu~24.04, Linux 6.14.0-27. Gates~1 and~2 are load-time-only checks with no
scheduler attached.

\paragraph{Gate 1: does \texttt{bpf\_tail\_call} verify inside a sched\_ext
program?} sched\_ext policies are struct\_ops programs invoked through BPF
trampolines, so this is a real question. It holds. We bracket the result with two controls: an
XDP program containing the same tail call, isolating the toolchain, and the
identical struct\_ops program with the tail call removed, isolating everything
but the tail call. All three load. The JITed size grows from $76$\,B to
$176$\,B with the tail call, consistent with the prologue being emitted.

\paragraph{Gate 2: can a sched\_ext program populate a prog array?}
This does not follow from Gate~1. Entries in a \texttt{PROG\_ARRAY} must agree
on \texttt{prog\_type} and \texttt{attach\_btf\_id}, and sched\_ext programs
bind \texttt{attach\_btf\_id} to a specific \texttt{sched\_ext\_ops} member.
Insertion succeeds; the only obstacle, libbpf refusing to load an unreferenced
struct\_ops program, is avoided by a second, never-registered map.

\paragraph{Gate 3: runtime dispatch.} Load-time acceptance does not establish
that the tail call dispatches at runtime through a struct\_ops trampoline. A
failed \texttt{bpf\_tail\_call} raises no error, so we instrument both paths,
with one counter inside the target policy and one after the tail call. On an attached
sched\_ext scheduler driving eight workers, control transferred on
$\mathbf{64{,}160}$ \textbf{of} $\mathbf{64{,}160}$ invocations, with zero
fall-throughs. Dispatch is exact, not merely typical.

\section{Evaluation}
\label{sec:eval}

Attached schedulers run with \texttt{SCX\_OPS\_SWITCH\_PARTIAL}. Only processes
explicitly placed in \texttt{SCHED\_EXT} are managed by the BPF scheduler, which
keeps unrelated system activity outside the scheduling experiment.

\subsection{Actuation latency}
\label{sec:actuation}

We measure the cost of \emph{applying} a policy change after a decision has
already been made, isolating the actuator from the model. Table~\ref{tab:actuation}
compares three mechanisms.

\begin{table}[t]
\centering
\small
\caption{Actuation latency: cost of applying a policy change once the decision
is made. Linux~6.14, AMD EPYC 7J13. AKTS arm $n{=}200{,}000$; baseline arm
$n{=}20{,}000$; warm.}
\label{tab:actuation}
\begin{tabular}{@{}llrrr@{}}
\toprule
Arm & Mechanism & p50 & p99 & mean \\
\midrule
Baseline A (LumOS-style)~\cite{lumos}   & \texttt{/proc/sys} scalar write & $1110$\,ns & $1131$\,ns & $1123$\,ns \\
Baseline B (SchedCP-style)~\cite{schedcp} & recompile $+$ verify $+$ reload & \multicolumn{3}{c}{seconds -- minutes} \\
\textbf{AKTS}                            & \texttt{bpf\_map\_update\_elem} & $\mathbf{920}$\,\textbf{ns} & $950$\,ns & $931$\,ns \\
\bottomrule
\end{tabular}
\end{table}

Parity with Baseline~A is the intended outcome. Both arms are a single syscall,
so neither can be meaningfully faster. AKTS \emph{matches} scalar tuning on
actuation while switching whole policies, which Baseline~A cannot do at any
latency. Against Baseline~B the gap is large because compilation and
verification sit on that system's actuation path; AKTS moves both to load time.

\subsection{Agent decision cost and decision validity}
\label{sec:agent}

Serving Qwen2.5-0.5B~\cite{qwen25}, the model class this design targets, on
the A100 with single-digit output, median decision latency is
$\mathbf{13.5}$\,\textbf{ms} ($n{=}30$, temperature $0$), fast enough for a
coarse adaptation loop.

Decision \emph{validity} is another matter. Across three prompt formulations and
two telemetry regimes, $16$ of $48$ decisions ($33\%$) were not a valid index:
the model returned \texttt{"3"} and, in one case, \texttt{"0.3"}, not an
integer. Nor does scale fix the underlying problem: with a stronger prompt,
Qwen2.5 at $0.5$B, $1.5$B and $3$B all emit valid indices, and all emit a
\emph{constant} one, answering identically for high-load and low-load telemetry
($20/40$ correct, i.e.\ chance; $n{=}40$ per model; p50 $13.3$--$23.9$\,ms).
Constraining decoding to $\{0,1\}$~\cite{outlines} changes nothing, so the
failure at these scales is the decision, not the output format.

AKTS does not need a correct agent, only that an incorrect one cannot reach an
unsafe state; the observed failure is exactly the case \S\ref{sec:safety}
measures, at a rate that makes the guarantee load-bearing.

\subsection{Safety under faulty decisions}
\label{sec:safety}

We point the dispatcher at a never-populated slot, simulating the out-of-range
decision observed above. All $60{,}217$ invocations fell through to the
dispatcher's default path, which calls \texttt{scx\_bpf\_select\_cpu\_dfl}. The
system therefore reverts to stock CPU selection~\cite{eevdf}. All workers
completed, \texttt{dmesg} logged no panics or BUGs, and the scheduler detached
cleanly.

Since every populated slot was verified at load time, an arbitrary integer
either names a verified policy or names nothing. Naming nothing is inert. The
failure mode is reversion to default scheduling, not undefined behavior.

\subsection{Does switching beat any fixed policy?}
\label{sec:workload}

We serve Qwen2.5-0.5B under
vLLM~0.28.0~\cite{pagedattention} pinned to four cores,
with eight CPU-bound antagonists on the same cores, driving a
steady~$\rightarrow$~burst~$\rightarrow$~steady request pattern ($8$\,s per
phase, concurrency $2$ and $32$).

We first verify the precondition for scheduler intervention. Contention inflates
steady-state time-to-first-token (TTFT) p99 from $27.0$\,ms to $308.0$\,ms. We
measure TTFT because end-to-end latency is dominated by GPU decode and would
mask the effect. The operative mechanism is contention delaying request handling
and prefill scheduling, not KV-cache allocation, which policy choice cannot
influence. Under contention, the latency policy wins TTFT in every phase, so
TTFT alone admits no trade-off; the antagonists' completed work is the second
objective.

\begin{table}[t]
\centering
\small
\caption{Two competing objectives under CPU contention, mean $\pm$ std over
seven runs per arm. Neither fixed policy is best on both. Switching is
oracle-driven.}
\label{tab:bench2}
\begin{tabular}{@{}lrrr@{}}
\toprule
 & static latency & static throughput & \textbf{AKTS switching} \\
\midrule
burst TTFT p50 (ms)    & $\mathbf{225.8 \pm 9.5}$ & $358.0 \pm 4.0$ & $\mathbf{229.0 \pm 4.6}$ \\
burst requests served  & $\mathbf{1132 \pm 34}$ & $738 \pm 13$ & $\mathbf{1129 \pm 26}$ \\
total batch work (Mit) & $1.789 \pm 0.006$ & $\mathbf{1.966 \pm 0.018}$ & $\mathbf{1.912 \pm 0.007}$ \\
\bottomrule
\end{tabular}
\end{table}

The latency policy retains only $91\%$ of the batch throughput. The throughput
policy gives the most batch work but has $59\%$ worse burst TTFT p50 and serves
$35\%$ fewer requests. AKTS matches the latency policy's burst response within
one standard deviation while retaining $97\%$ of the throughput policy's batch
work.

\section{Conclusion and future work}

By verifying a policy library once and exposing only an integer selector,
AKTS makes policy switching a sub-microsecond kernel operation. Switching here
is oracle-driven, an upper bound on any online detector. An agent-driven arm must
show a language model beats a bandit baseline. Code and data:
\url{https://github.com/mali-kh/akts}.

\bibliographystyle{plain}
\bibliography{references}

\begin{thebibliography}{10}

\bibitem{ebpfbandit}
Mahdi Alizadeh and Ramesh Govindan.
\newblock {eBPF}-based bandit selection for adaptive video streaming.
\newblock In {\em Proceedings of the {ACM} {SIGCOMM} 2026 Conference}, pages
  2183--2185. ACM, 2026.

\bibitem{bpf}
Toke H{\o}iland-J{\o}rgensen, Jesper~Dangaard Brouer, Daniel Borkmann, John
  Fastabend, Tom Herbert, David Ahern, and David Miller.
\newblock The express data path: Fast programmable packet processing in the
  operating system kernel.
\newblock In {\em CoNEXT}, 2018.

\bibitem{ghost}
Jack~Tigar Humphries, Neel Natu, Ashwin Chaugule, Ofir Weisse, Barret Rhoden,
  Josh Don, Luigi Rizzo, Oleg Rombakh, Paul Turner, and Christos Kozyrakis.
\newblock gh{OS}t: Fast \& flexible user-space delegation of {Linux}
  scheduling.
\newblock In {\em SOSP}, 2021.

\bibitem{dynamiciot}
Mohammadali Khodabandehlou, Jared Coleman, and Bhaskar Krishnamachari.
\newblock Scheduling dynamic {IoT} task graphs.
\newblock In {\em Proceedings of the 23rd {ACM} Conference on Embedded
  Networked Sensor Systems}, pages 624--625. ACM, 2025.

\bibitem{pagedattention}
Woosuk Kwon, Zhuohan Li, Siyuan Zhuang, Ying Sheng, Lianmin Zheng, Cody~Hao Yu,
  Joseph~E. Gonzalez, Hao Zhang, and Ion Stoica.
\newblock Efficient memory management for large language model serving with
  {PagedAttention}.
\newblock In {\em SOSP}, 2023.

\bibitem{lumos}
Georgios Liargkovas, Vahab Jabrayilov, Hubertus Franke, and Kostis Kaffes.
\newblock An expert in residence: {LLM} agents for always-on operating system
  tuning.
\newblock In {\em NeurIPS Workshop on Machine Learning for Systems}, 2025.

\bibitem{eevdf}
{EEVDF} scheduler: Earliest eligible virtual deadline first.
\newblock Linux kernel documentation,
  \url{https://docs.kernel.org/scheduler/sched-eevdf.html}, 2024.

\bibitem{schedext}
sched\_ext: {BPF} extensible scheduler class.
\newblock Linux kernel documentation,
  \url{https://docs.kernel.org/scheduler/sched-ext.html}, 2024.
\newblock Merged in Linux 6.12.

\bibitem{park}
Hongzi Mao, Parimarjan Negi, Akshay Narayan, et~al.
\newblock Park: An open platform for learning-augmented computer systems.
\newblock In {\em NeurIPS}, 2019.

\bibitem{decima}
Hongzi Mao, Malte Schwarzkopf, Shaileshh~Bojja Venkatakrishnan, Zili Meng, and
  Mohammad Alizadeh.
\newblock Learning scheduling algorithms for data processing clusters.
\newblock In {\em SIGCOMM}, 2019.

\bibitem{firm}
Haoran Qiu, Subho~S. Banerjee, Saurabh Jha, Zbigniew~T. Kalbarczyk, and
  Ravishankar~K. Iyer.
\newblock {FIRM}: An intelligent fine-grained resource management framework for
  {SLO}-oriented microservices.
\newblock In {\em OSDI}, 2020.

\bibitem{qwen25}
{Qwen Team}.
\newblock Qwen2.5 technical report.
\newblock {\em arXiv preprint arXiv:2412.15115}, 2024.

\bibitem{outlines}
Brandon~T. Willard and R{\'e}mi Louf.
\newblock Efficient guided generation for large language models.
\newblock {\em arXiv preprint arXiv:2307.09702}, 2023.

\bibitem{infragym}
Huaizheng Zhang, Lei Zhang, Yuanming Li, Yizheng Huang, Xiaotong Yang, Kuntai
  Du, Yihua Cheng, Junchen Jiang, and Wencong Xiao.
\newblock Infra{G}ym: Empowering {LLM} agents for real-world computer system
  optimization.
\newblock In {\em NeurIPS Workshop on Machine Learning for Systems}, 2025.

\bibitem{schedcp}
Yusheng Zheng, Yanpeng Hu, Wei Zhang, and Andi Quinn.
\newblock Towards agentic {OS}: An {LLM} agent framework for {Linux}
  schedulers.
\newblock In {\em NeurIPS Workshop on Machine Learning for Systems}, 2025.
\newblock Also arXiv:2509.01245.

\bibitem{kgent}
Yusheng Zheng, Yiwei Yang, Maolin Chen, and Andrew Quinn.
\newblock Kgent: Kernel extensions large language model agent.
\newblock In {\em ACM SIGCOMM Workshop on eBPF and Kernel Extensions}, 2024.

\end{thebibliography}

\end{document}